\documentclass{article}
\usepackage{smc2024}
\usepackage[caption=false, font=footnotesize]{subfig}
\usepackage{paralist}
\usepackage[figure,table]{hypcap}
\usepackage{caption}
\usepackage[english]{babel}

\usepackage{alphabeta}
\usepackage[LFE,LAE,LGR,T2A,T1]{fontenc}

\def\papertitle{Guitar Chord Diagram Suggestion for Western Popular Music}

\author[1]{\mbox{\firstname{Alexandre}\lastname{D'Hooge}
\email{alexandre.dhooge@algomus.fr}
\orcid{0000-0003-1634-3406}}}
\author[2]{\mbox{\firstname{Louis}\lastname{Bigo}}
\orcid{0000-0002-9865-2861}}
\author[1]{\mbox{\firstname{Ken}\lastname{Déguernel}
\orcid{0000-0001-7919-3463}}}
\author[3]{\mbox{\firstname{Nicolas}\lastname{Martin}}}

\affil[1]{\institution{Univ. Lille, CNRS, Centrale Lille, UMR 9189 CRIStAL}\postcode{F-59000}\city{Lille}\country{France}\affiliationtype{University}}
\affil[2]{\institution{Univ. Bordeaux, CNRS, Bordeaux INP, LaBRI, UMR 5800}\postcode{F-33400}\city{ Talence}\country{France}\affiliationtype{University}}
\affil[3]{\institution{Arobas Music}\city{Lille}\country{France}\affiliationtype{Company}}
\completesetup

\usepackage{algpseudocode}
\usepackage{svg}
\usepackage{multirow}
\usepackage{array}
\usepackage{siunitx}
\newcolumntype{C}[1]{>{\centering\let\newline\\\arraybackslash\hspace{0pt}}m{#1}}

\newcommand{\fingering}[1]{\texttt{#1}}
\newcommand{\cl}[1][]{\ensuremath{\mathrm{\ell}_{#1}}}
\newcommand{\cd}[1][]{\ensuremath{\mathrm{d}_{#1}}}

\newcommand{\chord}[1]{\emph{#1}}

\newcommand{\PP}{\ensuremath{\mathrm{P}_\mathrm{P}}}
\newcommand{\RP}{\ensuremath{\mathrm{R}_\mathrm{P}}}
\newcommand{\FP}{\ensuremath{\mathrm{F1}_\mathrm{P}}}
\newcommand{\PSF}{\ensuremath{\mathrm{P}_\mathrm{SF}}}
\newcommand{\RSF}{\ensuremath{\mathrm{R}_\mathrm{SF}}}
\newcommand{\FSF}{\ensuremath{\mathrm{F1}_\mathrm{SF}}}

\newcommand{\msbtracks}{520~}
\newcommand{\dadagptracks}{2766~}
\newcommand{\dadagpCTs}{31321~}
\newcommand{\msbCTs}{7365~}

\title{\papertitle}
\begin{document}
\capstartfalse
\maketitle
\capstarttrue

\begin{abstract}
Chord diagrams are used by guitar players 
to show where and how to play a chord on the fretboard. 
They are useful to beginners learning chords 
or for sharing the hand positions required to play a song.
However, the diagrams presented on guitar learning tools
are usually selected from an existing database of common positions
and rarely represent the actual positions used by performers.
In this paper, we propose a tool which suggests a chord diagram given a
chord label,
taking into account the diagram of the previous chord.
Based on statistical analysis of the DadaGP and mySongBook datasets, 
we show that some guitar chord diagrams are over-represented in Western popular music
and that some chords can be played in up to 108 different ways.
We argue that taking the previous position into account as context
can improve the variety and the quality of chord diagram suggestion, 
and compare this approach with a model 
taking only the current chord label into account.
We show that adding previous context 
improves the F1-score on this task by up to 32\% 
and reduces the propensity of the model to suggest standard open chords.
We also define the notion of texture in the context of chord diagrams and
show through a variety of metrics that our model improves texture
consistency
with the previous diagram.
\end{abstract}

\section{Introduction}
\label{sec:introduction}

\begin{figure}[ht]
    \centering
    \includegraphics[trim={0 35px 0 0},clip,width=.5\columnwidth]{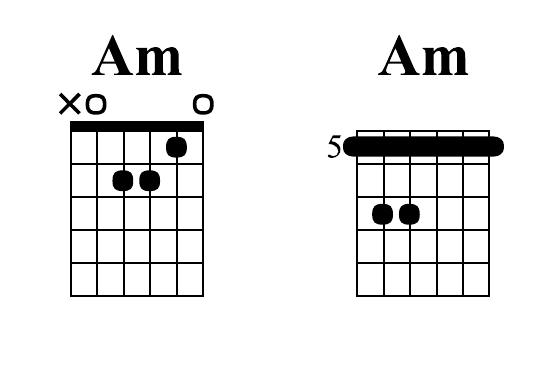}
    \caption{Two guitar chord diagrams for an A minor chord. 
    }
    \label{fig:chord-diagram}
\end{figure}

On the guitar, a chord can be played in multiple ways,
each position having its own pitch, timbral and biomechanical specificities.
\emph{Chord diagrams} can be used to represent 
the position at which a chord is played. 
They can be notated with a graphical representation as shown in \autoref{fig:chord-diagram},
or in text form (compatible with ASCII tablatures) 
with a number indicating which fret is played on each string.
For instance, the chord diagrams from Figure 1 
can be annotated, respectively,  as \fingering{x.0.2.2.1.0}, and \fingering{5.7.7.5.5.5}.
The former contains both an open (\texttt{0}) and a muted (\texttt{x}) string
and is the most common shape, used by beginners for its simplicity and in pop
music for the resonance of open strings. The latter is a \textit{barré} chord
and is harder to play, but can be shifted along the fretboard to
play other minor chords without changing hand shapes.

\vspace{-4\baselineskip}
We propose the task of chord diagram suggestion to assist composition. 
Chord diagram suggestion research has so far been focusing on playability 
\cite{sawayamaSystem2006, wortmanCombinoChord2021, vasquezQuantifying2023}
with limited ability to take into account
the relationship between consecutive chords.
We can see similar problems in the suggestions 
of the most popular guitar learning services, 
which only provide catalogs of standard chord diagrams.
While such an approach provides agency to the user, 
it might also drive beginners towards using
the same diagrams over and over again.
In this paper, we propose a model that suggests a chord diagram, 
given a chord label and the previous notated diagram.
The contributions of this work are as follows: 
(i) a context-aware approach for guitar chord diagram suggestion; 
(ii) a set of metrics to assess performance in this task 
and characterize texture for guitar chord diagrams; 
(iii) an illustrative application of this task for rhythm guitar tablature continuation; 
(iv) openly released code and data for all of the above.
The rest of this paper is organized as follows:
In \autoref{sec:sota}, we present related work on
symbolic music for guitar in Western popular music repertoire.
We then present our proposed approach and models in \autoref{sec:methodo}.
In \autoref{sec:data}, we discuss in details the datasets and the data augmentation procedure we used.
After discussing evaluation metrics, 
we finally share our results in \autoref{sec:results},
before showing how it could be used for rhythm guitar 
continuation in \autoref{sec:discussion}.

\section{Related Works}
\label{sec:sota}

A large part of research on guitar focuses on
tablatures as a notation system,
especially when studying Western popular music.
From the audio realm, tablatures are often
the output medium for guitar transcription 
\cite{yazawaAutomaticTranscriptionGuitar2014, humphreyMusicAudioChord2014, wigginsGuitar2019},
but they can even be used as a source of additional data to improve
automatic transcription
\cite{cwitkowitzDataDrivenMethodologyConsidering2022a, zangSynthTab2024}. 
In a field related to audio transcription,
chord recognition from audio is an important task in Music Information Retrieval
\cite{mcvicarAutomaticChordEstimation2014},
as attested by its recurring occurrence in 
MIREX\footnote{\url{https://www.music-ir.org/mirex/wiki/2021:Audio_Chord_Estimation}}and
ongoing research in the neighboring field of multipitch estimation 
\cite{bittnerLightweightInstrumentAgnosticModel2022a}. Chord recognition from audio was also specifically
studied in the case of guitar, exploiting knowledge of the instrument to improve
algorithms dedicated to it 
\cite{barbanchoAutomatic2012, humphreyMusicAudioChord2014, yazawaAutomaticTranscriptionGuitar2014}.
Once or when available, tablatures can also be both the input and the output, like for symbolic music
generation (symbolic referring here to the use of a notation format,
like sheet music or tablatures). This topic was already studied using Markov chains
\cite{mcvicarAutoGuitarTab2015} and benefited from the advance of deep neural 
networks \cite{chenAutomatic2020}. The release of the dataset DadaGP in 2021
\cite{sarmentoDadaGP2021} also had a significant impact in the field, fostering
all kinds of research on guitar tablature generation 
\cite{adkinsLooperGP2023,sarmentoGTRCTRL2023,sarmentoShredGP2023,lothProgGP2023, cuiMoodLoopGP2024}.
To assist composition, research has also been conducted on how to \textit{jazzify} chords \cite{chenChordJazzificationLearning2020} or generate style-conditioned
chord sequences \cite{dalmazzoChordinator2024}. Tablatures and
guitar chord research is also intertwined with the field of guitar fingering
research, as deriving tablatures from audio -- or scores -- requires
choosing string/fret combinations to play the notes, which will ultimately
be influenced by the fingers used to press the strings and the accompanying
biomechanical constraints. This problem has been tackled by a graph-search approach
\cite{lombardoGuitarFingeringMusic2005} or HMM-like models 
\cite{horiInputOutput2013, horiMinimax2016, horiThreeLevel2021} and can then
be used for tasks like automatic arrangement \cite{arigaSong2Guitar2017}.
 When it comes to guitar
chord diagrams however, they are mostly studied with playability considerations
in mind \cite{sawayamaSystem2006, wortmanCombinoChord2021,
vasquezQuantifying2023} or in a musicological analysis of their use \cite{cournutWhatAreMost2021}. In this paper, we consider chord diagram
suggestion as an assisted composition task, to help beginner guitar players choose
how to play a new chord in a preexisting sequence, or assist
composers in creating guitar accompaniment parts.

\section{Methodology}
\label{sec:methodo}
\subsection{Proposed Model}

\begin{figure}[ht]
    \includegraphics[width=\columnwidth,trim={0 1cm 0 0}]{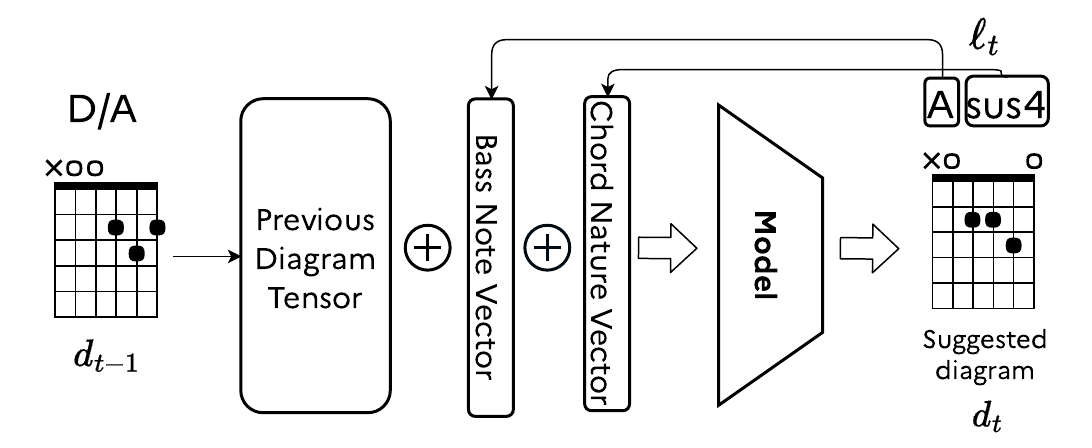}
    \vspace{.5\baselineskip}
    \caption{Summary of the proposed approach.}
    \label{fig:model-diagram}
\end{figure}

In this work, we suggest a diagram $\cd[t]$ for a chord,
based on its label $\cl[t]$ and the previous diagram $\cd[t-1]$ 
(\autoref{fig:model-diagram}). 

The task consists in finding the diagram 
which probability is highest in the provided context.
To do so, we convert the chord labels and diagrams 
into vectors and train a neural network on chord pairs.
For each chord label, we first extract its bass note,
which might be different from its root note in case of inverted chords.
This bass note information is then converted 
into a \emph{one-hot} vector of size 12
where enharmonic equivalents of the twelve-tone equal temperament are merged together 
(the \textit{Bass Note Vector}).
Then, the pitch-class content of the chord is converted
into a \emph{many-hot} vector of size 12 (the \emph{Chord Nature Vector}). 

\begin{figure}
    \centering
    \includegraphics[width=\columnwidth]{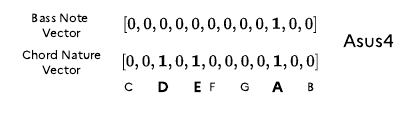}
    \caption{Computed vectors for an \chord{Asus4} chord, played with the diagram shown \autoref{fig:model-diagram}.}
    \label{fig:vectors}
\end{figure}

An example of such vectors for the \chord{Asus4} chord from \autoref{fig:model-diagram} is shown \autoref{fig:vectors}.

The diagrams are converted into many-hot arrays of size $[6, 25+1]$, 
each row accounting for a string (6 on a standard guitar) 
and its 25 frets (counting the open string as a zeroth fret).
An additional per-string coefficient is added 
to account for muted strings. 
This ensures that all vectors always have a non-zero value for a string, 
which ultimately permits normalizing predictions and using them as 
probabilities. 
In summary, input data contains 180 values for $\cd[t-1]$ -- the previous diagram --
and $\cl[t]$ --  the new chord label, from which the bass note and chord nature vectors
are computed.
As a result, the model outputs 156 probabilities, 26 per string, for $\cd[t]$.

To evaluate to which extent adding the context information 
improves chord diagram suggestion,
we define the baseline as the same model,
but removing information about the previous diagram.
In that case, the baseline model takes 
only 24 values as input, but still returns 156.

\subsection{Implementation Details}

We use a fully connected neural network 
as an architecture.
Hyperparameters were tuned -- through manual exploration --
to maximize performance 
of both the baseline and the proposed model, 
ending up with no hidden layer for the baseline and one hidden layer
of size 150 for the full model. 
Performance is evaluated with a
Binary Cross-Entropy loss on the output, 
the model using a Sigmoid activation function, 
and string-wise normalization of the predictions 
(to enforce a single prediction per string).
Training uses the Adam optimizer with $(\beta_1, \beta_2) = (0.9, 0.999)$,
and a learning rate $\lambda=0.001$. 
The training is stopped whenever the validation loss 
does not improve by at least $\delta=0.001$ for two consecutive epochs.
Full training of the model can be done in a few minutes on a standard laptop CPU.
The Python implementation is openly available, along
with a demonstration website, at \url{algomus.fr/code}.


\section{Data}
\label{sec:data}

\subsection{Corpora}
This study is conducted on two corpora of guitar tablatures 
containing chord diagram information. 
The first one is DadaGP \cite{sarmentoDadaGP2021}, 
a community-based corpus of more than 25000 songs 
spanning a wide variety of styles, though mostly focusing on rock and metal.
The second one is the proprietary dataset \textit{mySongBook}\footnote{\url{mysongbook.com}}(MSB),
which consists of about 2500 songs professionally transcribed.

The \texttt{.gp} file format of these datasets 
allows including chord diagrams with the tablature data.
We only use tracks which contain chord diagram data, 
reducing the datasets to \dadagptracks and \msbtracks tracks 
for DadaGP and MSB respectively. 
From these tracks, we extract chord pairs that occur within a 2 bars interval
and finally keep one occurrence of each unique transition 
$(\cl[t-1],\cd[t-1]) \rightarrow (\cl[t], \cd[t])$ per track.
By doing so, we acknowledge the fact that a chord transition
can be more common in a given repertoire and occur in many songs, 
but we remove duplicate transitions that are inherent to the
repetitive nature of Western popular music.
This processing step leaves us with \dadagpCTs and \msbCTs transitions
for DadaGP and MSB, respectively.

\subsection{Statistical Analyses}
\label{ssec:stats}

In this subsection, we provide detailed statistics of the chords used
in the datasets. This analysis aims at emphasizing the bias towards
more common diagrams and key signatures, due to guitar affordance.
Furthermore, because of the considered repertoire, and the large
amount of pop/rock/metal tracks, some chord natures are more common than 
others. However, we also want to underline that this bias is not at the
expense of variety, but results in highly unbalanced datasets.
\smallskip

\begin{figure}[ht]
    \centering
    \includegraphics[width=\columnwidth]{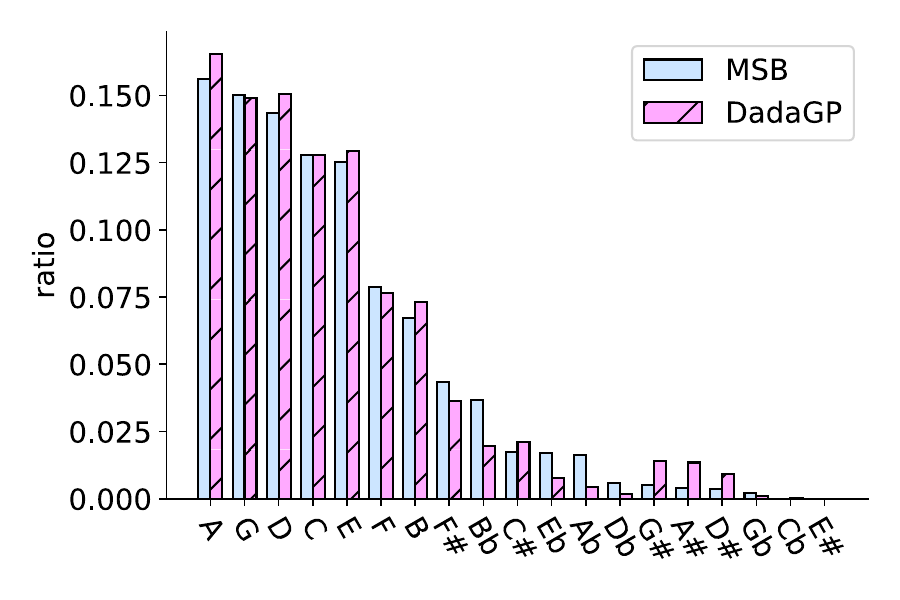}
    \vspace{-.7cm}
    \caption{Distribution of the root notes of chords in both datasets.}
    \label{fig:root-notes}
\end{figure}

A first dimension to analyze on chords is on what root note they
are built.
Observation shows that the distributions of root notes are similar between datasets 
(\autoref{fig:root-notes}), with more than half the chords
having \textit{A, G, D, C} or \textit{E} as root. 

\begin{figure}[ht]
    \centering
    \includegraphics[width=\columnwidth]{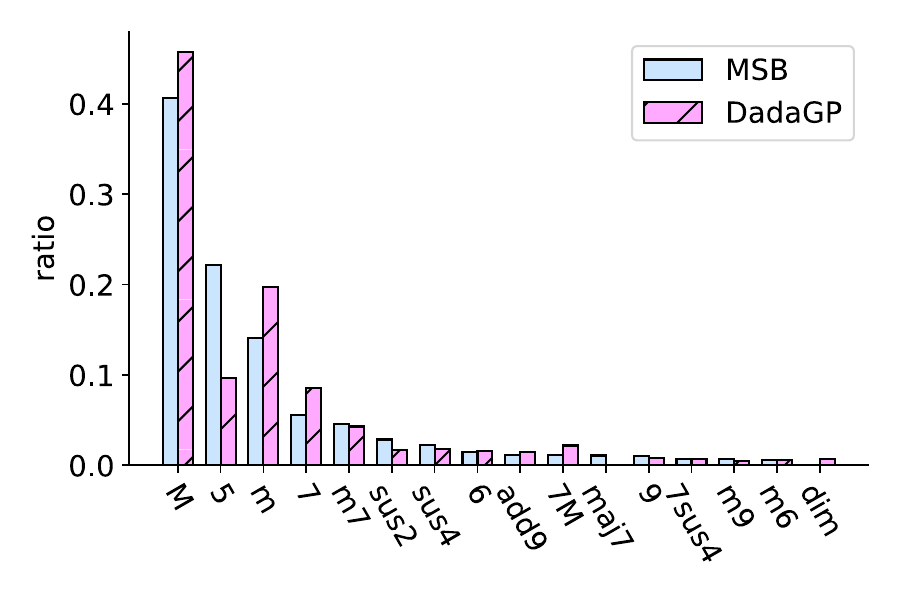}
    \vspace{-.7cm}
    \caption{Distribution of the 15 most used chord natures in each dataset.}
    \label{fig:chord-nature}
\end{figure}

In both datasets, major chords (M) are the most common 
(\autoref{fig:chord-nature}), 
followed by minor chords (m) and \textit{power chords} (5).
The tail of the distribution then includes more complex chords, 
like seventh, suspended or added-tone chords.
The distribution of chord natures is in fact 
highly unbalanced towards the three first classes, 
making the suggestion of diagrams for less common chords a bigger challenge.
Besides, this plot shows that notation between files can vary, 
MSB containing lots of \chord{7M} and \chord{maj7} 
labels that are both representing major seventh chords.

\begin{figure}[t]
    \centering
    \includegraphics[width=.4\columnwidth]{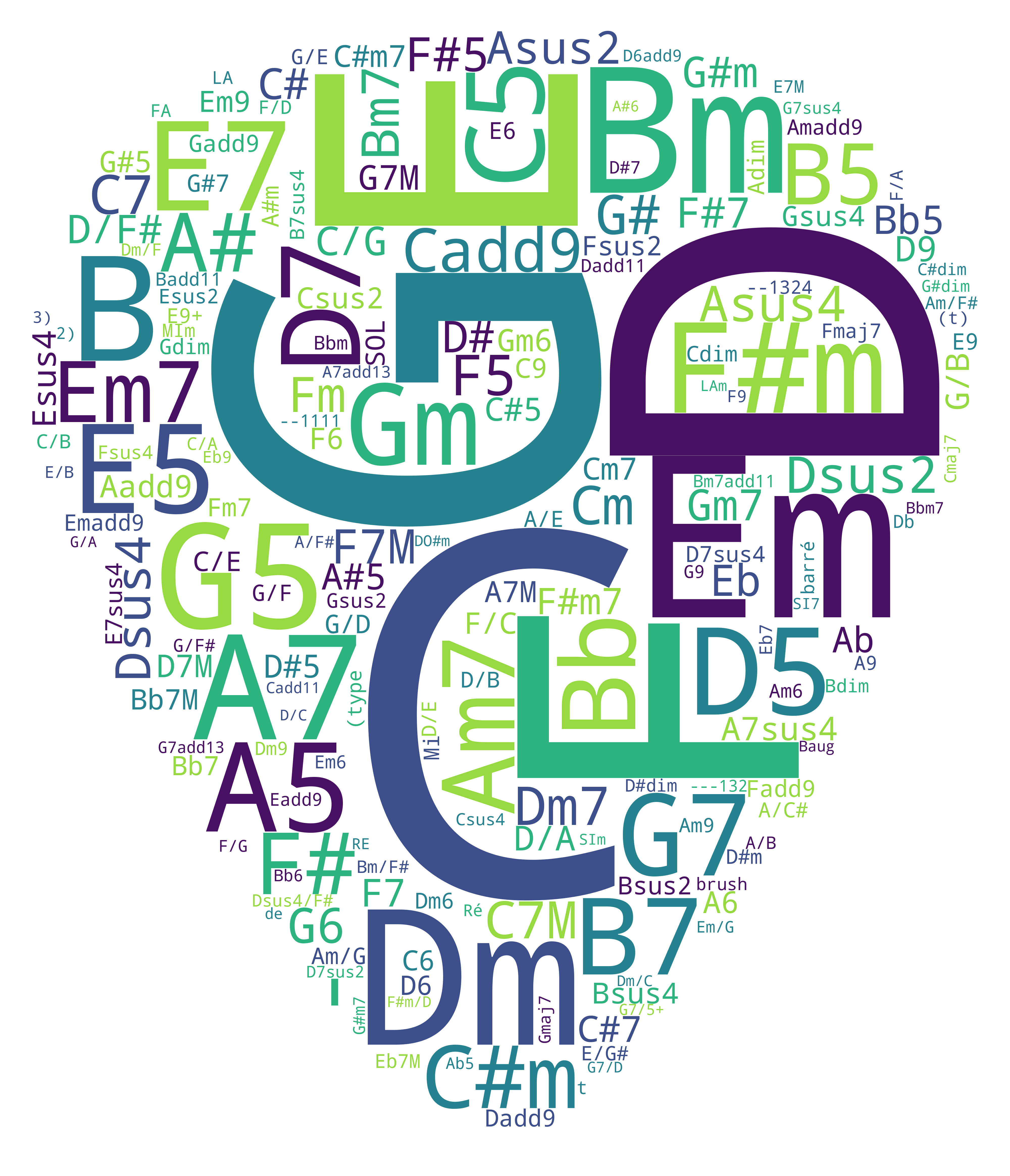}
    \includegraphics[width=.4\columnwidth]{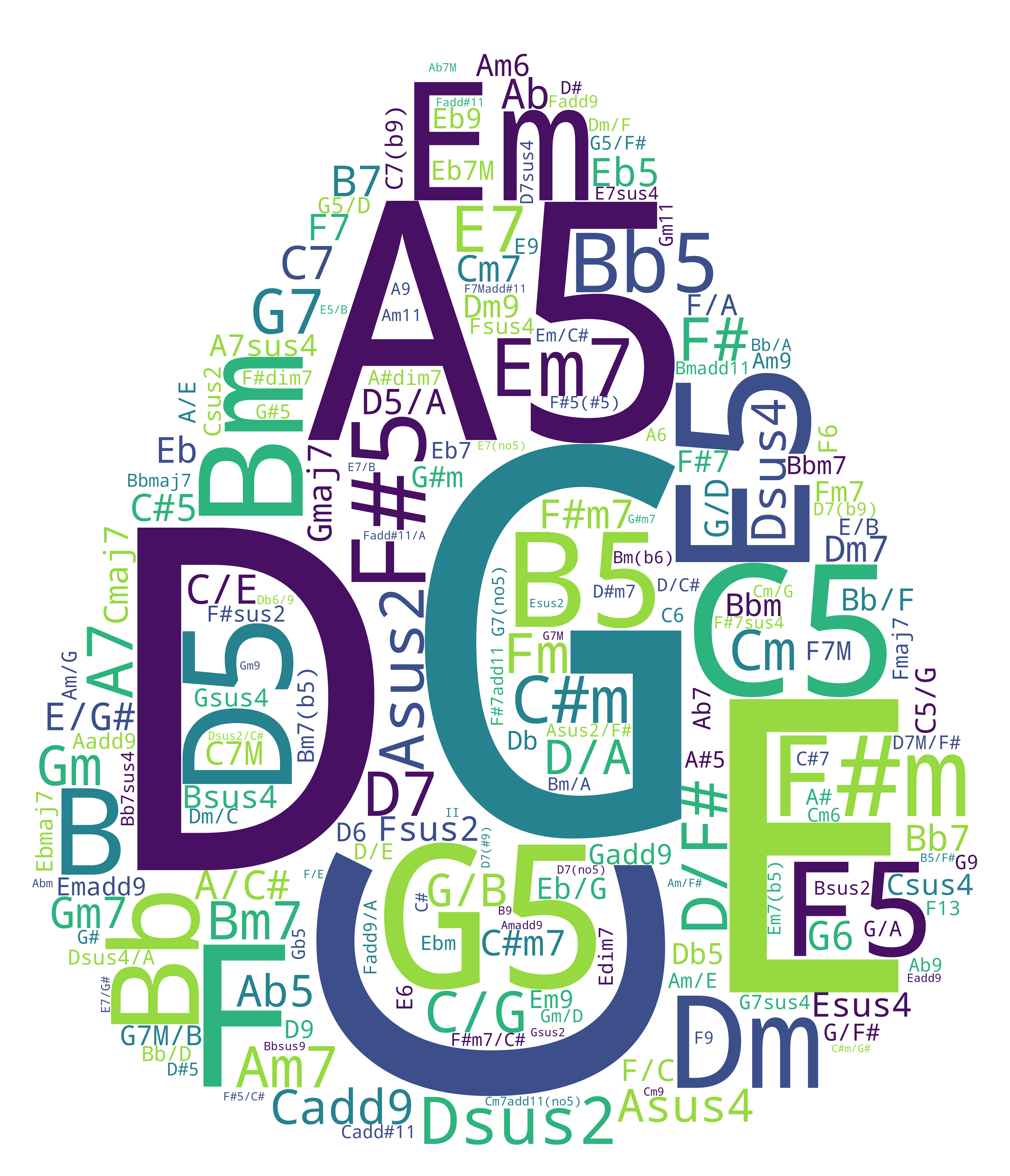}
    \caption{Wordcloud representation of chord labels in DadaGP (left) and
    MSB (right)}
    \label{fig:wordclouds}
\end{figure}

Wordcloud representations of the chord labels are also provided (\autoref{fig:wordclouds})
to give an overview of the combined root notes and chord natures. This representation
further confirms the unbalanced nature of the datasets but also illustrates 
their differences
in the relative importance of chords. For instance, \emph{power chords}
 are more prominent in MSB than in DadaGP.

While chord labels are already numerous and varied, 
even more variety comes with the different diagrams that can be used for each label. 
One can get a sense of it through the median number of diagrams per label: 11.5 for MSB and 24 for DadaGP. 
This observation illustrates the number of possibilities when suggesting a diagram for a chord,
while also showing how a bigger dataset comes with more variety (which could be considered noise).
This
variety and noisiness of data is particularly noticeable on the most common
chords, G Major having 33 and 108 diagrams in MSB and DadaGP, respectively.
The variety of diagrams observed comes mostly from the fact that transcribers might
write some strings as muted instead of open or played, which can yield different diagrams
for the same hand position on the fretboard. The staggering amount of diagrams in DadaGP
comes both from the same issue, worsened by the size of the corpus, and occasional incorrect labeling.
For instance, some diagrams notated as a G Major can actually represent a 
\chord{G/B} or a \chord{G5}.


All diagrams observed for each chord label are 
available, along with the extracted chord pairs, on \url{algomus.fr/data}.

\subsection{Data Augmentation Strategy}
\label{ssec:data-aug}

The previous statistical analyses showed that 
a wide variety of diagrams and labels are represented in both datasets. 
However, more common chords are over-represented
and might prevent the model to suggest less used diagrams. 
We want to improve the robustness of the proposed approach and reduce its bias
towards key signatures more frequent in the datasets so that the model
can suggest diagrams even for rare tonalities and chord nature.
To do so, we apply the augmentation technique from \cite{mcvicarAutoGuitarTab2015}. 
For each chord pair, both diagrams are shifted one
fret down and the chord labels transposed one semitone down accordingly, until
one of the diagrams contains an open string. Similarly, chord pairs are also shifted one
fret up until reaching the 15th fret. This maximum is chosen based on the highest
diagrams observed in data and to prevent the model from suggesting chords on
higher frets, which are used far less in rhythm guitar.
Using this augmentation strategy makes the training sets more than three times larger.

\section{Experiments}
\label{sec:results}

\subsection{Metrics}

Like most tasks on probability estimation, 
we can evaluate the performance of the model through 
its Precision (P), Recall (R) and F1-score. 
However, in the case of this musical task, 
it could be argued that the so-called ground-truth 
is not perfect nor the only acceptable answer. 
For instance, a G major chord can be played with the following diagrams:
\fingering{3.2.0.0.0.3} and \fingering{3.2.0.0.3.3} either repeating
 the B or the D, and they might be considered equivalent by guitar players.
For this reason, we propose hereafter several automated metrics 
to account for the different errors 
that can be encountered on diagram prediction.
\smallskip

\textbf{Pitch Metrics} 
Similarly to \cite{wigginsGuitar2019}, 
we want to measure to which extent the model 
suggests diagrams that contain all the expected notes for a given chord label.
For this purpose, we compute for a chord diagram $d$
the set of pitch classes it contains and compare it 
with the pitch classes associated with the chord label $\ell$.\\
From these sets, we can compute Pitch Precision ($\PP$), 
Pitch Recall ($\RP$) and a Pitch F1-score ($\FP$).



\smallskip

\textbf{Tablature Metrics} 
Drawing insights from \cite{wigginsGuitar2019}, 
we also define metrics to measure
how similar the model suggestions are to the reference diagrams,
when comparing them on the fretboard. 
Similarly to the pitch metrics, 
we compute the set of string/fret (SF) pairs on the predicted diagram 
and compare them with those of the expected diagram.
We can use those sets to compute String/Fret Precision ($\PSF$), 
String/Fret Recall ($\RSF$) and the corresponding F1-score ($\FSF$).
\smallskip

\textbf{Detect Unplayable Diagrams} 
To evaluate how often the model can return an unplayable diagram, 
we implement a playability metric inspired by \cite{wortmanCombinoChord2021}. 
In this paper, the authors define an \textit{anatomical score} 
to evaluate the ease of playing a chord. 
They use a custom fitness function that penalizes 
chords containing uncomfortable stretches between fingers 
based on an optimal fingering computed beforehand. 
This metric allows us to detect unplayable diagrams when obtaining a low anatomical score, 
like in chords spanning over 5 frets or more.
We settled for a threshold of $t=0.2$ after manual analysis 
of anatomical scores and playability of the model's suggestions.

\textbf{Ease of Transition} 
While it is necessary that the suggested diagrams are playable, 
playability of chord sequences also depends on the chord transitions involved. 
To assess the ease of the transitions suggested,
we also implement the chord change (CC) metric 
proposed in \cite{yazawaAutomaticTranscriptionGuitar2014}.
This metric analyzes the transition between
the diagrams $d_1, d_2$ through two movements: 
 \emph{wrist movement} $m_w$,
which is obtained as the absolute difference of the index finger fret 
in the two chords; and the \emph{fingers movement} $m_f$,
defined as the Manhattan distance between the positions of each finger. 
The final value is inversely proportional to these movements measures 
and is highest (1) when the chord change is easy ($m \ll 1$) 
and close to 0 when the transition is very hard ($m \gg 1$).

\subsection{Results}
\label{ssec:results}
For the experiments, we apply a 60-20-20 
train, validation, test split to both datasets
and average the results over four different splits
(\autoref{tab:results}). 
For fair comparison with the baseline, 
chord pairs which have identical $\cl[t]$ and $\cd[t]$ (but different $\cd[t-1]$)
are considered duplicates and skipped during testing. 
The data augmentation strategy presented \autoref{ssec:data-aug} 
is used for the full model but not for the baseline 
as it decreased its performance. 
This observation suggests that the results
of the baseline should be taken with care, 
as prediction quality might decrease significantly on 
data not represented in the training set. 
The first observation from \autoref{tab:results} 
is that the proposed model surpasses the baseline on standard, 
pitch and string/fret F1-scores.
However, both implementations perform well on pitch metrics, 
showing that the proposed diagrams contain overall the expected pitch content.
There is also a significant improvement over the baseline 
on string/fret metrics, suggesting that information
from the previous diagram helps the model choose the correct fretboard area.

\begin{table}[ht]
    \centering
    \begin{tabular}{|cl|ccc|}
	    \cline{3-5}
	    \multicolumn{1}{c}{}&    & F1    & $\FP$ & $\FSF$ \\
                \hline
	    \multirow{2}{*}{\rotatebox[origin=center]{90}{MSB}}    & \multicolumn{1}{|l|}{Baseline} &  0.40$\pm$.01  & 0.87$\pm$.01 & 0.46$\pm$.04\\
	    & \multicolumn{1}{|l|}{Full Model} &  0.72$\pm$.02  & 0.90$\pm$.01 & 0.67$\pm$.02\\
    \hline \hline 
	    \multirow{2}{*}{\rotatebox[origin=center]{90}{DGP}}    
	    & \multicolumn{1}{|l|}{Baseline}   & 0.38$\pm$.01  & 0.88$\pm$.00 & 0.45$\pm$.02\\
	    & \multicolumn{1}{|l|}{Full Model} & 0.63$\pm$.01  & 0.88$\pm$.01 & 0.60$\pm$.02\\
    \hline
    \end{tabular}
    \caption{Results of the baseline and the full model on MSB (top) and DadaGP (bottom). Precision and Recall measures were omitted for clarity.}
    \label{tab:results}
\end{table}

Unplayability and transition ease of the ground-truth 
and the models' prediction are shown in \autoref{tab:data-stats}.
It can be observed that the proposed model suggests 
unplayable diagrams 15\% of the time, 
which is moderately more than the baseline,
with a clearer gap on DadaGP.
As a reference, 3-4\% of the ground-truth diagrams are deemed unplayable 
-- most of them because the metric does not recognize 
the \textit{barré} technique
with other fingers than the index. 
However, the proposed diagrams permit slightly easier transitions than the baseline, 
which is probably again due to the context information
that keeps the suggested diagram in the same fretboard area, 
thus limiting wrist and fingers movements. Nonetheless,
even our model with context information has a 10 percentage points difference
with the transition ease measured in the datasets, showing that the transitions
are still too complicated.

\begin{table}[t]
    \centering
    \begin{tabular}{|cl|C{0.2\columnwidth}C{0.2\columnwidth}|}
	    \cline{3-4}
	    \multicolumn{1}{c}{}&        & Unplayable diagrams & Ease of Transition  \\
         \hline
\multirow{3}{*}{\rotatebox[origin=center]{90}{MSB}}    
	    &\multicolumn{1}{|l|}{Baseline}   & 0.12$\pm$.02                & 0.15$\pm$.02                \\
	    &\multicolumn{1}{|l|}{Full Model} & 0.15$\pm$.03                & 0.21$\pm$.03                \\
    \cline{2-4}
	    &\multicolumn{1}{|l|}{Test Set}& 0.03$\pm$.02                & 0.31$\pm$.02                \\
    \hline \hline
\multirow{3}{*}{\rotatebox[origin=center]{90}{DadaGP}}    
	    &\multicolumn{1}{|l|}{Baseline}   & 0.09$\pm$.01 & 0.13$\pm$.01                \\
	    &\multicolumn{1}{|l|}{Full Model} & 0.15$\pm$.01 & 0.19$\pm$.01                \\
    \cline{2-4}
	    &\multicolumn{1}{|l|}{Test Set}   & 0.04$\pm$.01 & 0.29$\pm$.01                \\
    \hline 
    \end{tabular}
    \caption{Playability-related metrics on MSB (top) and DadaGP (bottom).}
    \label{tab:data-stats}
\end{table}

Overall, one can note that performance is similar on MSB and DadaGP. 
We still decided to share results on both datasets 
because we deemed relevant the fact that the proposed approach 
improves diagram suggestion even when training on noisier data. 
It also shows that a larger amount of chord pairs with
an unbalanced distribution does not significantly increase the bias 
towards common diagrams.

\subsection{Diagram Texture Consistency}

We want to evaluate here if the \textit{texture} change observed when using a suggested diagram
is similar to the one from the reference. To measure this texture, we implement
some of the sound quality measures of \cite{wortmanCombinoChord2021} and variations
of them. More precisely, for each diagram, we extract: the ratio of open strings; 
the ratio of muted strings; the string centroid; and the ratio of unique notes
\textit{i.e.} counting only once notes that are repeated on several octaves.
We also compute the difference of these metrics from one chord to the next to
assess how consistent they are through a transition.
The results are reported in \autoref{tab:texture}.

\begin{table*}[ht]
    \centering
    \begin{tabular}{|c|cc|c||cc|c|}
        \cline{2-7}
        \multicolumn{1}{c|}{} & \multicolumn{3}{c||}{DadaGP} & \multicolumn{3}{c|}{MSB} \\
        \cline{2-7}
        \multicolumn{1}{c|}{} & BL   & Full & Data & BL   & Full & Data \\
         \hline
    Muted notes     & 0.15 {\scriptsize$\pm$ .01} & 0.27 {\scriptsize$\pm$ .01} & 0.28 {\scriptsize$\pm$ .01} & 0.32 {\scriptsize$\pm$ .02} & 0.32 {\scriptsize$\pm$ .01} & 0.33 {\scriptsize$\pm$ .01} \\  
    $\delta(.)$     & 0.20 {\scriptsize$\pm$ .01} & 0.08 {\scriptsize$\pm$ .00} & 0.10 {\scriptsize$\pm$ .01} & 0.18 {\scriptsize$\pm$ .02} & 0.08 {\scriptsize$\pm$ .01} & 0.08 {\scriptsize$\pm$ .01} \\  
         \hline
    Open Strings    & 0.25 {\scriptsize$\pm$ .01} & 0.19 {\scriptsize$\pm$ .02} & 0.20 {\scriptsize$\pm$ .01} & 0.20 {\scriptsize$\pm$ .01} & 0.17 {\scriptsize$\pm$ .02} & 0.20 {\scriptsize$\pm$ .02} \\  
    $\delta(.)$     & 0.22 {\scriptsize$\pm$ .01} & 0.15 {\scriptsize$\pm$ .01} & 0.17 {\scriptsize$\pm$ .00} & 0.19 {\scriptsize$\pm$ .01} & 0.15 {\scriptsize$\pm$ .01} & 0.15 {\scriptsize$\pm$ .01} \\  
         \hline
    String Centroid & 0.53 {\scriptsize$\pm$ .00} & 0.54 {\scriptsize$\pm$ .01} & 0.55 {\scriptsize$\pm$ .00} & 0.60 {\scriptsize$\pm$ .01} & 0.59 {\scriptsize$\pm$ .01} & 0.59 {\scriptsize$\pm$ .01} \\  
    $\delta(.)$     & 0.10 {\scriptsize$\pm$ .00} & 0.05 {\scriptsize$\pm$ .00} & 0.06 {\scriptsize$\pm$ .00} & 0.11 {\scriptsize$\pm$ .01} & 0.05 {\scriptsize$\pm$ .01} & 0.06 {\scriptsize$\pm$ .01} \\  
         \hline
    Unique Notes    & 0.61 {\scriptsize$\pm$ .00} & 0.71 {\scriptsize$\pm$ .00} & 0.80 {\scriptsize$\pm$ .01} & 0.68 {\scriptsize$\pm$ .01} & 0.72 {\scriptsize$\pm$ .01} & 0.80 {\scriptsize$\pm$ .01} \\  
    $\delta(.)$     & 0.21 {\scriptsize$\pm$ .01} & 0.13 {\scriptsize$\pm$ .00} & 0.12 {\scriptsize$\pm$ .00} & 0.20 {\scriptsize$\pm$ .02} & 0.13 {\scriptsize$\pm$ .01} & 0.11 {\scriptsize$\pm$ .01} \\  
         \hline
    \end{tabular}
    \caption{Texture metrics for chord suggestions of the proposed model (Full), the
    baseline (BL) and the corresponding test set (data). $\delta(.)$ denotes
    the absolute difference of the previous metric between the two chords of 
    a transition.}
    \label{tab:texture}
\end{table*}

All values go from 0 to 1. 
A first observation is that the performances of the full models trained and tested on DadaGP and
MSB are similar. However, this experiment does show slight differences in the datasets, on
the amount of muted notes and the string centroids in particular. 
The biggest difference is the improvement of the baseline's performance on
individual chord metrics for MSB, which could be due to the fact that the dataset is smaller and less varied.
From
the $\delta$ values for both datasets, it appears that the metrics are rather
consistent from one chord to the next. We can also observe that the baseline trained on DadaGP tends to play
too many notes on each chord (lower ratio of muted notes) probably by repeating
them on different octaves (lower ratio of unique notes). Finally, an encouraging result
is that the model using
context suggests diagrams with a texture similar to the ground-truth (lower
$\delta$ values). It should
however be noted that it also
tends to repeat more notes than necessary, while still missing some.

\section{Discussion}
\label{sec:discussion}

\subsection{Application to Rhythm Guitar Continuation}

\begin{figure}
    \centering
    \includegraphics[width=\columnwidth]{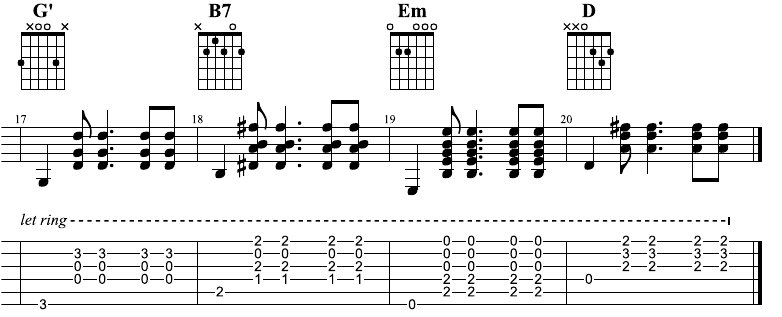}
    \includegraphics[width=\columnwidth]{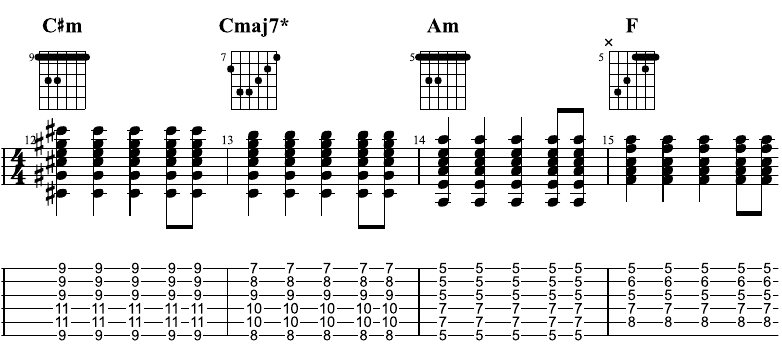}
    \includegraphics[width=\columnwidth]{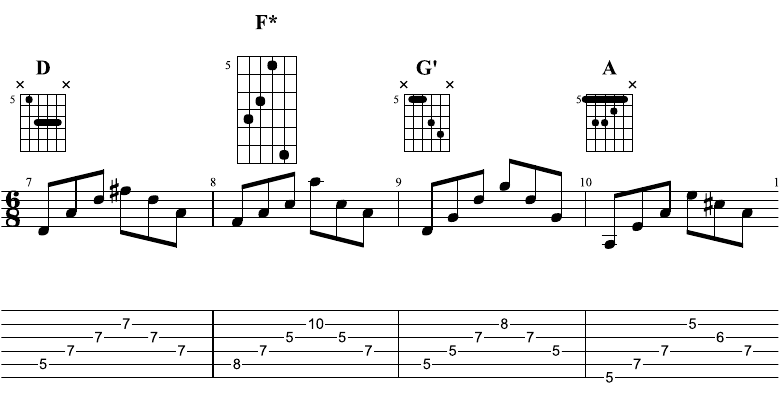}
    \vspace{-.5cm}
    \caption{Example of rhythm guitar continuation using the proposed system. The first diagram is chosen randomly, and the next ones
    are suggested from their label and the previous diagram.
    Chords with a star are unplayable according to our implemented
    metric, and chords with an apostrophe do not contain all the notes
    they should (G5 instead of G).}
    \label{fig:rhythm-gtr}
\end{figure}

As introduced in \cite{dhoogeRhythmGuitarTablature2023},
a chord diagram suggestion tool could be used
in a larger framework for rhythm guitar continuation. Provided with
a chord sequence and a tablature prompt, a guitar player might
be interested in generating the continuation of the prompt. The model
introduced in this paper could be used as a first step to choose chord
diagrams that are consistent with the prompt, before generating a tablature
showing how to strum the new chords. An example of this usage on three different chord sequences
is given \autoref{fig:rhythm-gtr}.

Chord label sequences are generated with the
tool from \cite{dalmazzoChordinator2024} and
the first diagram is sampled from the probability distribution of all diagrams
for the corresponding label. Subsequent diagrams are generated successively using our trained model. To illustrate potential tablature applications, strumming patterns were proposed
by the authors for the first bar, and then uniformly generalized to
following bars.
As discussed in \autoref{ssec:results}, the system sometimes suggests unplayable
diagrams and does not always include all the notes expected in a chord. However, we think that the sequences
are consistent in terms of texture and usually contain diagrams less common than
standard open chords and \textit{barré} chords. For instance, in the first sequence,
the B7 diagram contains an open string that would help maintain the sound
quality of the first chord, unlike the more common \textit{barré} shapes
that are \fingering{x.2.4.2.4.2} or \fingering{7.9.7.8.7.7}. In the second
sequence, though the second diagram is not playable (at least with common
guitar techniques), the model suggested keeping a similar \textit{barré} shape and
shifting it on the fretboard. However, it changes shape on the last F chord
which allows staying around the same frets as the previous Am chord.
Finally, the last sequence exhibits a similar behavior, all diagrams
starting at the 5th fret, even though the second diagram repeats the
A on the 10th fret, making the overall diagram very hard to play.

\subsection{Conclusions and Future Work}

In this paper, we have shown that chords used in the Western popular guitar repertoire
are varied but also highly unbalanced, with some common chords and diagrams being used much
more frequently than more complex ones.
From this observation, we proposed a new approach to suggest guitar chord diagrams for Western
popular music. We showed with several metrics that adding context information through 
the previous diagram improves the quality of the suggestions while also
maintaining a better consistency of texture between chords.
All experiments were conducted on two datasets, one proprietary and one public, to 
further guarantee the validity of the conclusions.
Finally, we also gave an example application of the proposed tool in rhythm
guitar continuation, where it could help beginners play more interesting and varied 
chords, or ease the process of writing accompaniment tracks.

\medskip

Studying this application in more details is a possibility for future work, 
as well as improving the current suggestion tool, for instance by conditioning
the suggestions on musical style. It would also be relevant to increase the
number of previous diagrams provided as context. This could allow seeing how far back
the system needs to look to maximize the quality of the suggestions, and
show when and how continuity in texture is broken. Besides, it is likely
that the suggestion system could be improved by musically relevant features on the context
like harmony or instrumentation, as other instruments affect greatly the diagrams chosen by guitar players.
Finally, while we studied texture consistency through several metrics, it would also
be interesting to check how the diagrams suggested affect the voices from one chord to the
other and whether the transitions respect some \textit{voice-leading} principles.

\newpage
\begin{acknowledgments}
    This work is partially funded under the ANR-TABASCO project:
    \href{https://anr.fr/Projet-ANR-22-CE38-0001}{ANR-22-CE38-0001}.
    The authors would like to thank the reviewers, as well as
    Baptiste Bacot, Louis Couturier, Dinh-Viet-Toan Le, and Mathieu Giraud for their
    feedback on this paper.
\end{acknowledgments} 

\bibliography{references}

\end{document}